# Superparamagnetic Superparticles for Magnetic Hyperthermia Therapy: Overcoming the Particle Size Limit

Supun B. Attanayake[1,#], Minh Dang Nguyen[2,#], Amit Chanda[1], Javier Alonso[3], Iñaki Orue[4], T. Randall Lee[2,*], Hariharan Srikanth[1], and Manh-Huong Phan[1,*]

[1]Department of Physics, University of South Florida, Tampa, FL 33620, USA

[2]Department of Chemistry and the Texas Center for Superconductivity, University of Houston, 4800 Calhoun Road, Houston, TX 77204-5003, USA

[3]Department of CITIMAC, Universidad de Cantabria, Santander 39005, Spain

[4]SGIker Medidas Magnéticas, Universidad del País Vasco, Leioa 48940, Spain

**Iron oxide (e.g., $Fe_3O_4$ or $\gamma\text{-}Fe_2O_3$) nanoparticles are promising candidates for a variety of biomedical applications ranging from magnetic hyperthermia therapy to drug delivery and bio-detection, due to their superparamagnetism, non-toxicity, and biodegradability. While particles of small size (below a critical size, ~20 nm) display superparamagnetic behavior at room temperature, these particles tend to penetrate highly sensitive areas of the body such as the Blood-Brain Barrier (BBB), leading to undesired effects. In addition, these particles possess a high probability of retention, which can lead to genotoxicity and biochemical toxicity. Increasing particle size is a means for addressing these problems but also suppresses the superparamagnetism. We have overcome this particle size limit by synthesizing unique polycrystalline iron oxide nanoparticles composed of multiple nanocrystals of 10 to 15 nm size while tuning particle size from 160 to 400 nm. These so-called *superparticles* preserve superparamagnetic characteristics and exhibit excellent hyperthermia responses. The specific absorption rates (SAR) exceed 250 W/g ($H_{AC}$ = 800 Oe, $f$ = 310 kHz) at a low concentration of 0.5 mg/mL, indicating their capability in cancer treatment with minimum dose. Our study underscores the potential of size-tunable** **polycrystalline iron oxide superparticles with superparamagnetic properties for advanced biomedical applications and sensing technologies.**

[#]Equal contributions to the work.

*Corresponding authors: phanm@usf.edu (M.H.P.); trlee@uh.edu (T.R.L.)

# Introduction

Nanotechnology, the science of material manipulation at the atomic level, continues to play a significant role in everyday applications.[1–5] Its impact spans across diverse fields, including electronics, computing, medicine, healthcare, energy, and environmental sectors. Traditionally, the dimensions, shapes, and compositions of materials have been central to nanotechnology research.[6,7] However, these parameters have become secondary to the control of phase-tunability and structural ordering, which allows for the fine-tuning of nanostructures for current and future applications.[8–13]

Magnetic hyperthermia therapy utilizes magnetic nanoparticles (magnetite $Fe_3O_4$ or maghemite $\gamma$-$Fe_2O_3$) to generate heat through the induction of an alternating current (AC) magnetic field.[14–18] When exposed to this AC magnetic field, magnetic nanoparticles can elevate the temperature of the body from the natural level to the effective hyperthermia range of 40 to 43°C, making them highly efficient for cancer treatment.[14,15] Hyperthermia therapy has been extensively explored as a secondary cancer treatment process with minimal adverse effects on the body while ensuring high efficacy in combating cancers via direct and indirect involvement.[14,19–22] The U.S. Food and Drug Administration (FDA) has approved certain iron oxide compositions and spherical nanostructures for the usability in many treatment processes.[23–26] The large-scale usage of such therapeutic technologies, due to their less invasive nature and minimal side effects, has further given hope in making them more effective and safer.[27] The application of nanostructures in highly sensitive areas of the body such as the remarkably sensitive blood-brain barrier (BBB) -- a semi-permeable membrane and the gateway to the central nervous system -- tends to carry an imminent high risk to the life of a patient as particles in the nano-level possess the caliber to pass through.[28–30] Therefore, it is essential to exploit magnetic nanoparticles with tunable sizes for such applications.

Unlike ferro/ferri-magnetism, superparamagnetism, which generally tends to establish in structures less than 20 nm (with zero remanent magnetization, $M_r \sim 0$ and coercivity, $H_c \sim 0$), is advantageous due to its easy manipulation, detectability, higher efficacy, and controllability, and is attractive for use in hyperthermia treatment processes, drug delivery and targeting, biodetection, and magnetic memory devices.[31–35] Traditionally, magnetic hyperthermia has primarily focused on the use of single-domain iron oxide nanoparticles with superparamagnetic (SPM) properties.[14,15,17,36] The use of SPM nanostructures has been further supported in hyperthermia treatments due to the facilitation of low agglomeration levels with increased dispersibility but without remanent magnetization, as the magnetization is easily flipped by the thermal energy, which exceeds the magnetic anisotropy energy.[37,38] However, due to their small size, these nanoparticles exhibit limited heating efficiency, arising from their low magnetization. Transitioning to larger nanoparticles with greater magnetization can enhance heat generation efficiency, but often leads to the emergence of ferromagnetic (FM) or ferrimagnetic (FiM) properties, which gives rise to magnetic clustering due to strong magnetic dipole interactions.[12,16] As noted above, the smaller size of nanoparticles also raises concerns about their ability to penetrate highly sensitive areas, which can give rise to potential issues related to genotoxicity and biochemical toxicity. In this context, the use of large and size-tunable polycrystalline nanoparticles with SPM properties offers an efficient solution to address the existing problems of single-domain magnetic nanoparticles.

The synergistic approach required to protect superparamagnetism while mitigating the risk in vulnerable areas has led us to scrutinize the structures more closely and propose a novel approach as illustrated in Fig. 1. Very large polycrystalline iron oxide nanoparticles, hereby named *Superparticles* (SUPAs), in the particle size range of 150 to 400 nm, can be synthesized, ruling out potential hazards while preserving the SPM nature owing to the 10 to 15 nm sizes of nanocrystals (crystallite size/grain size) within each SUPA that are smaller than the SPM size limit (~20 nm).[39–41] In this study, our emphasis was on exploring the magnetic hyperthermia characteristics of superparamagnetic SUPAs in a medium of 2% agar since it closely replicates conditions found in biological environments.[42–44] This novel approach has shown the potential of our SUPAs for the benefit of global cancer treatment since nearly 20 million new cancer patients and 10 million deaths are reported annually due to cancer and cancer-related effects.[45] The preservation of superparamagnetism in SUPAs with particle size tunability up to 400 nm also highlights them as excellent candidates for many other applications including the detection of single particle-based cells and targeted drug delivery.

**Novel Approach to Overcoming the Superparamagnetic Particle Size Limit**

Magnetic nanostructures, ranging from a few nanometers to a few hundred nanometers, offer the capacity to tune their magnetic properties based on size, shape, crystallinity, and composition.[11,17,46–48] The SPM behavior observed in FM and FiM materials implies a magnetization versus magnetic field response with zero coercivity ($H_c$) and zero remanence magnetization ($M_r$) above the blocking temperature ($T_B$).[46] Upon size reduction, $H_c$ varies with particle/grain size in a complex manner;[46,49] it increases with decreasing particle/grain size for multidomain magnetic systems (particle/grain size, $D > D_c$), reaches a maximum at a critical size ($D_c$) at which the system transforms from the multidomain (MD) state to the single domain (SD) state, declines to zero when particle/grain size decreases to $D_{SPM}$ (the SPM size limit), and remains zero below $D_{SPM}$ (Fig. 1a). In the MD regime, FM or FiM materials generally possess significant $H_c$ values since grain boundaries inhibit propagation of the magnetization reversal or magnetic domain boundary movement, which depends on the magnetic anisotropy and nucleation and growth of reverse domains.[50,51] As the structure morphs into an SD from an MD system, the resulting SD becomes increasingly difficult to align with the applied field, yielding the maximum $H_c$ at the $D_c$. With further decreases in particle/grain size, the thermal energy becomes increasingly important in governing the spin orientation above a certain critical temperature.[52]

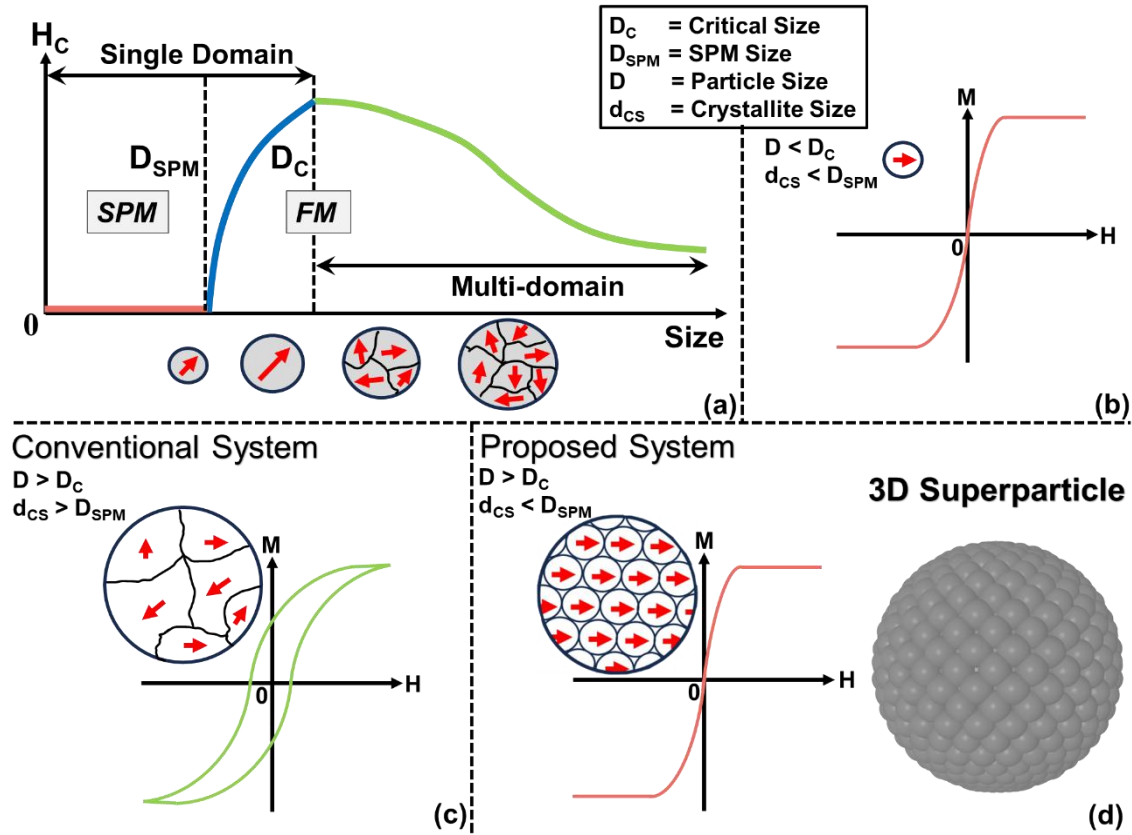

**Figure 1:** (a) The qualitative behavior of the coercivity and the domain boundaries in nanoparticles with size, (b) Hysteresis loop of superparamagnetic (SPM) particles with no coercivity ($H_c = 0$) and remanent magnetization ($M_r = 0$), (c) Hysteresis loop of ferro/ferrimagnetic (FM or FiM) multidomain particles with non-zero coercivity ($H_c \neq 0$) and remanent magnetization ($M_r \neq 0$), and (d) Hysteresis loop, domain states, and a 3D visualization of a superparticle showing SPM characteristics.

This mechanism enables the structures to show SPM behavior with no coercivity and no remanent magnetization (Fig. 1b), as the crystallite size ($d_{cs}$) of the structure falls below $D_{SPM}$. Unlike the cases of large SD FM or FiM particles or multi-domain FM or FiM particles with $d_{cs}$ > $D_{SPM}$ (Fig. 1c), we propose the creation of large polycrystalline particles composed of multi-nanocrystals with $d_{cs}$ < $D_{SPM}$, securing the SPM feature but with average particle size well above $D_c$ offering impassability through vital barriers (Fig. 1d). In this scenario, the average diameter of these SPM polycrystalline particles, reaching up to 400 nm, can far exceed that of the SD/MD FM or FiM particles that can span between 70 and 100 nm in diameter for magnetite, with variations based on their shape.[53,54]

## Results and Discussion

### *Structural characterization*

The structural evaluation of SUPA samples was initially carried out using a scanning electron microscope (SEM) to determine general morphology and uniformity. The sizes of the SUPA system enabled us to categorize them at varying particle sizes up to 400 nm, as summarized in Table 1. The uniformity of the structures in all five samples was consistent with no significant morphological deviations, as shown in Fig. 2(a,b) for samples S2 and S3, respectively. SEM images of samples S1, S4, and S5 are presented in Fig. S1.

Furthermore, transmission electron microscopy (TEM) was employed to evaluate the structural integrity, interparticle adjacency, and superficial crystallite distribution (Fig. 2c-e). Upon closer examination through subsequent qualitative analysis, we found that the average crystallite size ($d_{cs}$) remained consistently below 15 nm across all the structures. (see Table 1). The selected area electron diffraction (SAED) pattern, shown in the inset of Fig. 2(c) for sample S3, exhibits clear diffraction spots confirming the crystalline nature of SUPAs,[55] which reflects the long-range order inherent in these crystalline structures. The planes can be indexed back to magnetite composition, and the presence of dislocations in the structure has led to the streaks, and the partial rings illustrate the preferred orientations in the structures.[55] The structure and composition were further examined using X-ray diffraction (XRD). As shown in Fig. 2(f), all XRD data show well-defined peaks, revealing the good crystallinity of the SUPAs.

The position and relative intensities of these peaks match well with those of magnetite structure, confirming the presence of magnetite (Fe$_3$O$_4$) as a major phase in all the samples (Fe$_3$O$_4$ JCPDS card number: 01-088-0315). This hypothesis has been further confirmed by X-ray photoelectron spectroscopy (XPS) analysis (refer to the SI, Fig. S2). All samples exhibited the presence of the major phase magnetite (Fe$_3$O$_4$) and the minor phase maghemite (γ-Fe$_2$O$_3$).

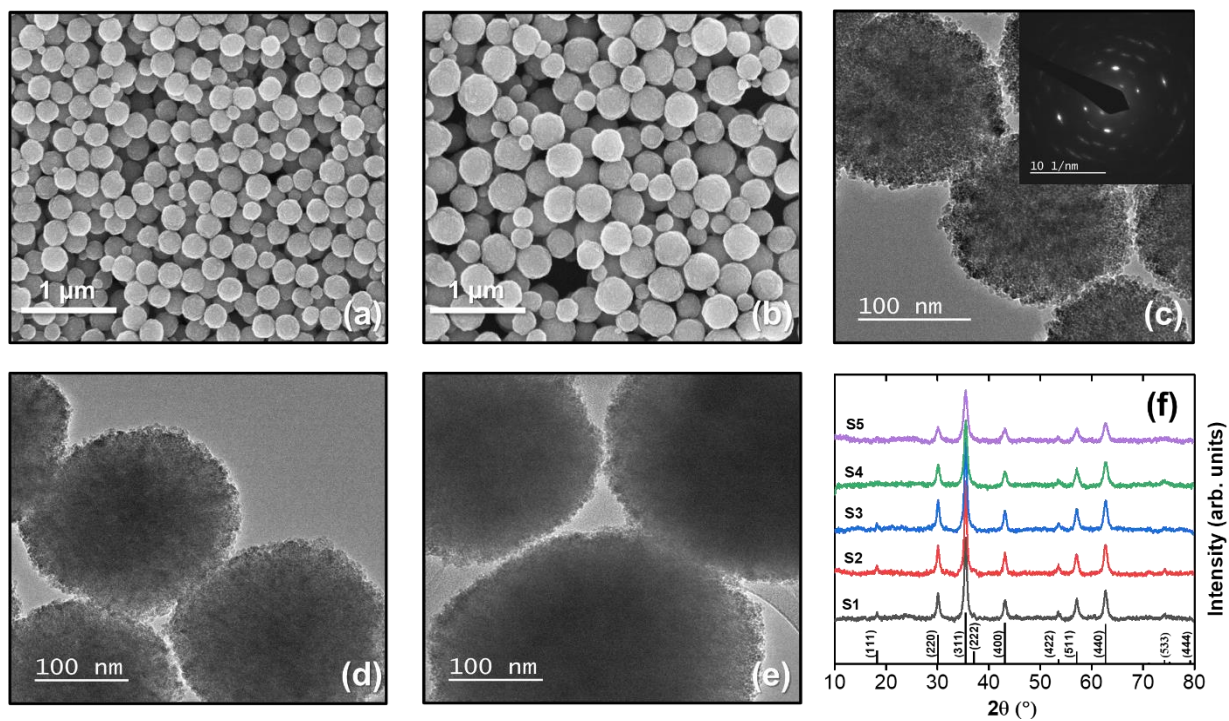

**Figure 2:** SEM images of samples (a) S2 and (b) S3; TEM images of samples (c) S2, (d) S3, and (e) S5. The inset of (c) shows a selected area electron diffraction (SAED) pattern of sample S2; (f) XRD patterns for all samples. The positions of the *hkl* reflections for bulk Fe$_3$O$_4$ are marked for reference.

**Table 1:** Particle size ($D$), crystalline size ($d_{cs}$), saturation magnetization ($M_s$), coercivity ($H_c$), and specific absorption rate (SAR) of the iron oxide superparticles. SAR values were measured in agar (0.5 mg/mL) at an AC field of 800 Oe and a frequency of 310 kHz.

| Sample | $D$ (nm) | $d_{cs}$ (nm) | SAR (W/g) | $M_s$ at 300 K (emu/g) | $H_c$ at 300 K (Oe) |
|---|---|---|---|---|---|
| S1 | 159 ± 11 | 12 | 286 | 66 | ~0 |
| S2 | 175 ± 11 | 15 | 253 | 68 | ~0 |
| S3 | 234 ± 15 | 12 | 251 | 60 | ~0 |
| S4 | 271 ± 31 | 10 | 252 | 62 | ~0 |
| S5 | 375 ± 37 | 10 | 241 | 61 | ~0 |

*Magnetic properties*

The magnetic properties of the SUPA samples were characterized by a Vibrating Sample Magnetometer (VSM) equipped within a Physical Property Measurement System (PPMS) from Quantum Design. The temperature dependence of the magnetization, $M(T)$ with zero-field cooled (ZFC) and field-cooled (FC) curves for all samples, was measured sequentially over 10 K ≤ $T$ ≤ 350 K range in the presence of a $\mu_0 H$ = 0.05 T applied field. Fig. 3(a) shows the $M(T)$ curves of a representative sample, S1. The $M(T)$ data of the remaining samples are displayed in Fig. S3. It is generally understood that FC magnetization curves tend to flatten out when reaching lower temperatures, which indicates that the sample has kept magnetization constant, while upward or downward shifting of the FC magnetization curve at low temperatures hints at interactions between particles that become more prominent as thermal energy decreases.[47] Fig. 3(a) shows that as the thermal energy decreases, the magnetization drops since the attractions between SUPAs have made it difficult to be aligned with the applied magnetic field. Similar behavior was observed across other samples as well (refer to the SI, Fig. S3).

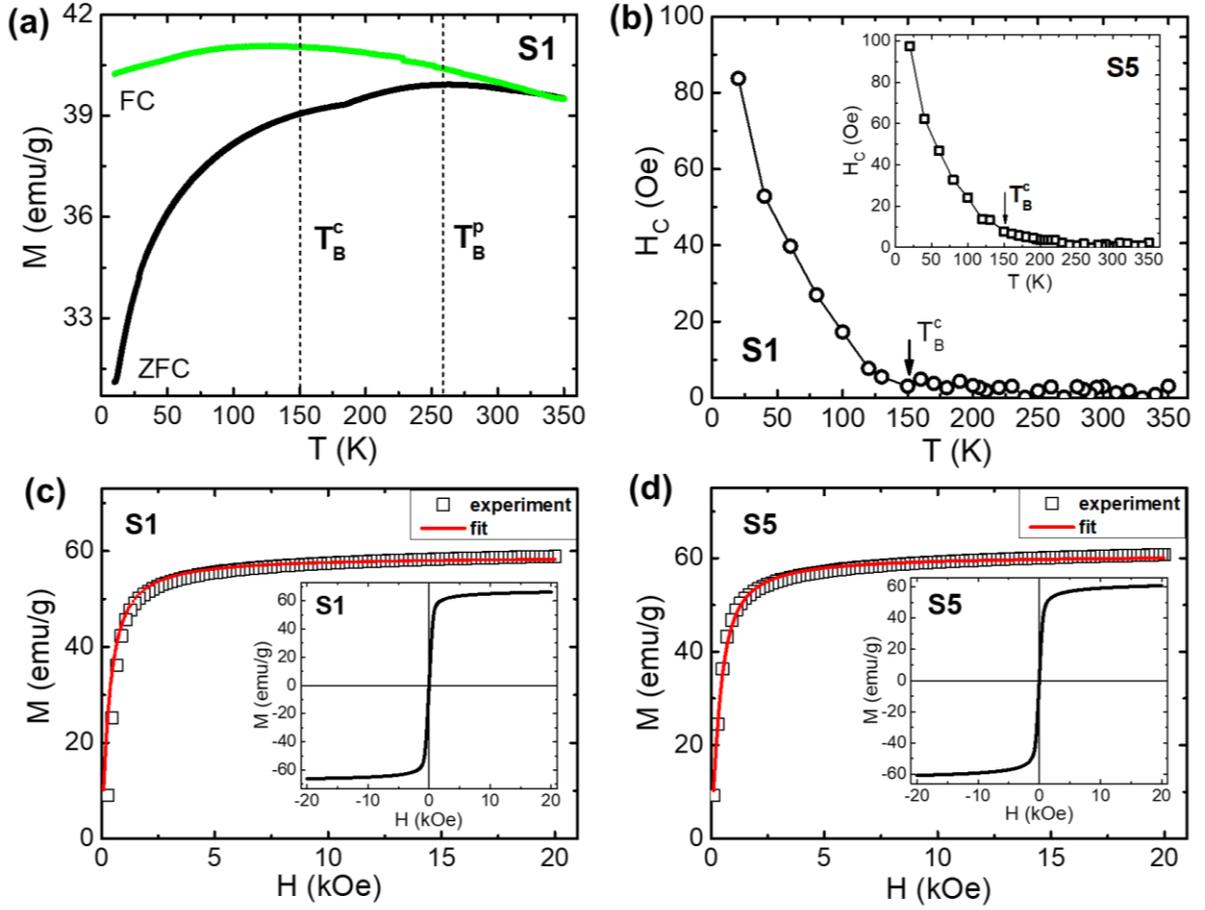

**Figure 3:** Magnetic characteristics of samples: (a) ZFC and FC $M(T)$ curves measured in an applied field of 0.05 T for sample S1; (b) Coercivity change with temperature for sample S1 and sample S5 (inset). Experimental and fitted $M$-$H$ curves at 300 K, with the $M(H)$ loops at 300 K shown in the insets, for (c) sample S1 and (d) sample S5. The fits were conducted utilizing the SPM model (Eq. (1)).

We recall that a peak or maximum observed in a ZFC $M(T)$ curve is commonly identified as the average blocking temperature ($T_B$) of a nanoparticle system above which $H_C$ tends to approach zero with a minimal temperature-dependent change.[56] As observed in Fig. 3a and Fig. S4, a conspicuous cusp or peak is evident in the FC magnetization curves at $T_B^p$ ~250 K across all SUPA samples. However, it should not be simply attributed to the average blocking temperature of SUPAs. Instead, it likely represents the onset temperature of magnetic ordering of the nanocrystals within each SUPA. Interestingly, this observation holds true regardless of particle size variations, ranging from 160 nm (S1) to 370 nm (S5). It is anticipated that all SUPA samples should manifest a SPM-like feature at $T > T_B^p$. The absence of magnetic hysteresis in the $M(H)$ loops observed for all SUPA samples has indeed confirmed the SPM behavior at room temperature (see insets of Fig. 3c,d, and Fig. S4). A detailed analysis of the $H_c$ vs. $T$ plots for all samples, as showcased in Fig. 3b and its inset for sample S1 ($D = 160$ nm) and sample S5 ($D = 370$ nm), revealed a notable increase in $H_c$ below $T_B^C \sim 150$ K, marking a clear transition from the SPM to FM or FiM (blocked) state. It is noteworthy that the significant decrease in magnetization ($M$) below ~150 K in the ZFC $M(T)$ curve (Fig. 3a) correlates with the pronounced increase in $H_c$ as the temperature drops below this threshold (Fig. 3b), suggesting the blocking of nanocrystals within each SUPA occurring below $T_B^C \sim 150$ K. Herein, $T_B^C$ refers to the average blocking temperature of nanocrystals encapsulated within SUPAs. Our findings underscore that the magnetic properties of SUPAs are primarily determined by the size of the nanocrystals composing each superparticle. The observation of broadened ZFC $M(T)$ curves across all SUPA samples (Fig 3a and Fig. S3) is notable, likely stemming from competing interactions among nanocrystals within each SUPA and among different SUPAs, as well as from the size distributions of nanocrystals and SUPAs.

To provide a deeper insight into the SPM behavior, we fitted the room temperature $M(H)$ loops for all samples using a well-known SPM model:[57]

$$M(H) = \int_0^\infty M_0 L\left(\frac{\mu_0 M_S V H}{k_B T}\right) f(D)\, dD \qquad (1)$$

where $V$ and $D$ are the volume and diameter of the nanocrystals, respectively; $M_0$ is the saturation magnetization reached in the experimental $M$–$H$ loops; $L(x) = \coth(x) - 1/x$ is the so-called Langevin Function; and $M_S$ is the theoretical saturation magnetization of magnetite (i.e. ~450 emu cm$^{-3}$ for bulk magnetite at 300 K).[58] The function $f(D)$ corresponds to the particle size distribution; in our case, we have assumed a log-normal size distribution, typical for nanoparticles, as shown below:

$$f(D) = \frac{1}{D\beta\sqrt{2\pi}} \cdot \exp\left(-\frac{(\ln D - \ln \alpha)^2}{2\beta^2}\right) \quad (2)$$

where $\alpha$ and $\beta$ are fitting parameters, and the other fitting parameter is $M_0$. From the fittings, we can estimate the average "magnetic diameter" $\overline{D}$ and standard deviation $\sigma$ of the nanoparticles:

$$\overline{D} = \alpha\exp\left(\beta^2/2\right), \quad \sigma^2 = \overline{D}^2\left(\frac{\overline{D}^2}{\alpha^2} - 1\right) \quad (3)$$

A good fit was obtained for all SUPA samples, as showcased in Figs. 3c and 3d for samples S1 and S5, respectively. The fitting results for other samples are also included in the SI (Fig. S5). The estimated average size and standard deviation are $\overline{D}$ = 9.4 nm and $\sigma$ = 0.1 nm for S1, and $\overline{D}$ = 9.3 nm and $\sigma$ = 0.9 nm for S5, respectively. These values are close to the ones obtained by TEM for the crystallites (see Table 1), confirming that at 300 K these nanocrystallites are the ones dominating the SPM-like behavior of the SUPAs. These findings are in agreement with our aforementioned interpretation of the ZFC-FC $M(T)$ curves. It is noteworthy that through the careful control of crystallite size to remain below the SPM threshold ($d_{cs} < S_{SPM}$), we have successfully engineered superparamagnetic SUPAs, offering the capability to tune particle sizes up to 400 nm. In other words, the SPM characteristics of SUPAs are primarily influenced by the size of the nanocrystals within each individual SUPA, rather than the size of the SUPA itself. While tuning the particle size between 160 and 400 nm, consistent values of $M_s$ were maintained across all SUPA samples (refer to Table 1). These consistent characteristics hold significant promise for diverse biomedical applications. We note further that the optimum crystallite size range for maintaining superparamagnetic properties while preserving high saturation magnetization is 10 to 15 nm. Crystallite sizes below 10 nm retain superparamagnetism but have reduced magnetization capability,[59] while sizes above 15 nm might exhibit ferromagnetic behavior.[12]

*Magnetic hyperthermia properties*

To explore the potential of superparamagnetic SUPAs for biomedical applications, we tested their magnetic hyperthermia responses. The heating efficiency of the SUPAs was evaluated in both water and 2% weight agar solution at 0.5 mg/mL and 1 mg/mL, which are comparatively lower concentration levels compared to those reported previously.[60–62] The agar solution restricted the physical movement of SUPAs, mimicking cell environments such as cell cytoplasm and extracellular matrix.[63,64] Fig. 4 displays the heating curves for all samples at concentrations of 1 mg/mL, and 0.5 mg/mL, subjected to AC fields of 600 Oe and 800 Oe at a constant frequency of 310 kHz. The data show that all the samples yielded consistent results, with sample S1 showing significantly higher heating capability across all concentrations. Notably, the heating capabilities of the samples for the hyperthermia treatment process were evaluated for a maximum of 15 minutes (900 s), which is lower than the typical treatment duration of 30 minutes to 1 hour.[65,66] The natural cooling observed after the removal of $H_{AC}$ indicated satisfactory retention of heat, which is a promising feature.

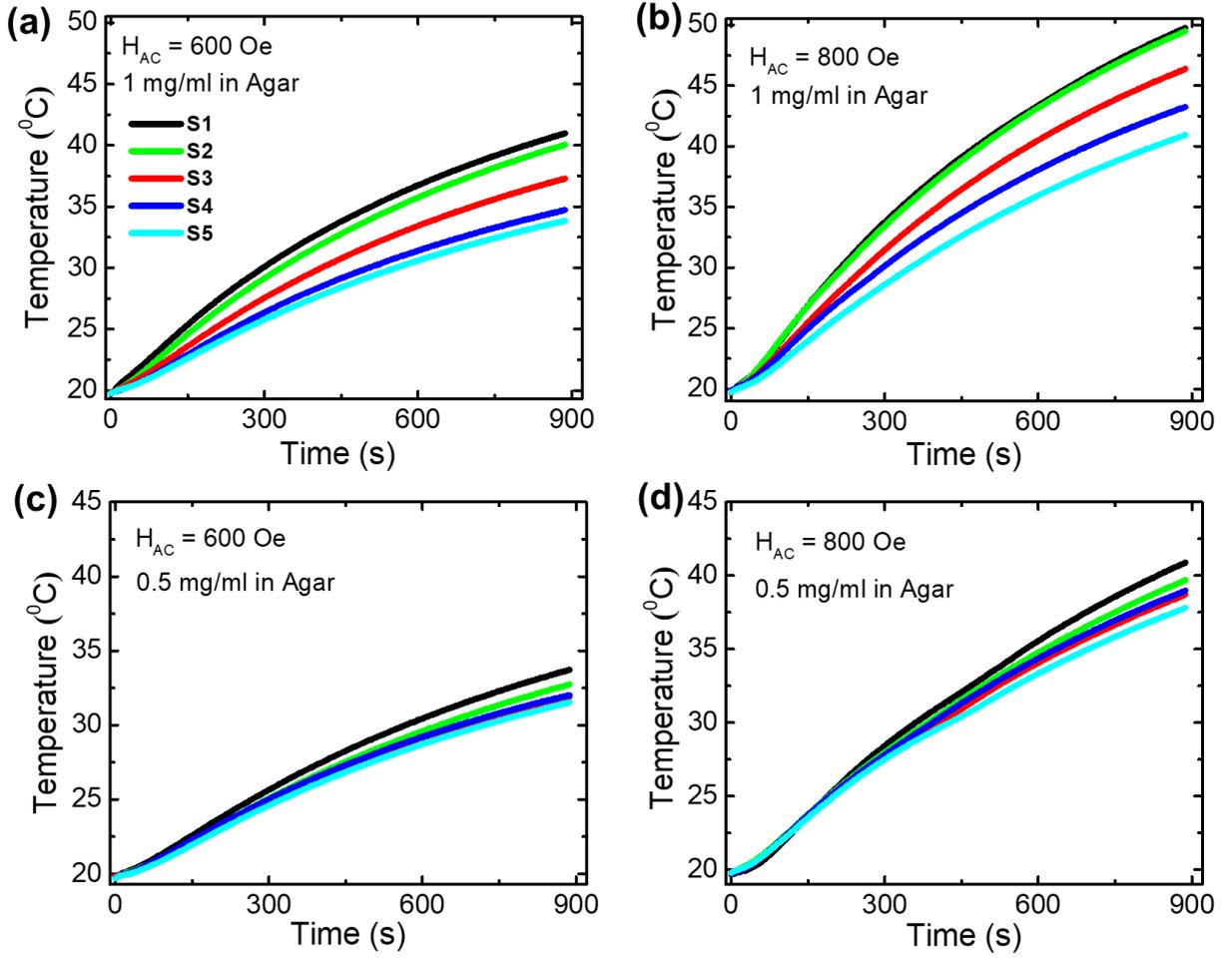

**Figure 4:** Temperature vs. time measurements or heating curves for all five samples at concentrations of 1 mg/mL and 0.5 mg/mL in the presence of 600 Oe and 800 Oe AC fields at a constant frequency of 310 kHz.

To quantify the heating efficiency of the samples, the Specific Absorption Rate (SAR), which is a measure of the absorption energy, usually determined from the initial rate of temperature rise, was evaluated.[67] In our computations utilizing Eq. (4), we derived the SAR values from the heating curves after a 60-second interval, as the SUPAs tend to stabilize over time:

$$SAR = \frac{\Delta T}{\Delta t} \times \frac{Cp}{\varphi} \qquad (4)$$

Here the $\Delta T/\Delta t$ represents the change in temperature with time, and $C_P$ represents the heat capacity of the liquid solvent (in this case, water with 4.186 J/g·K). The symbol $\varphi$ represents a unitless quantity: mass of magnetic material per unit mass of liquid solvent.

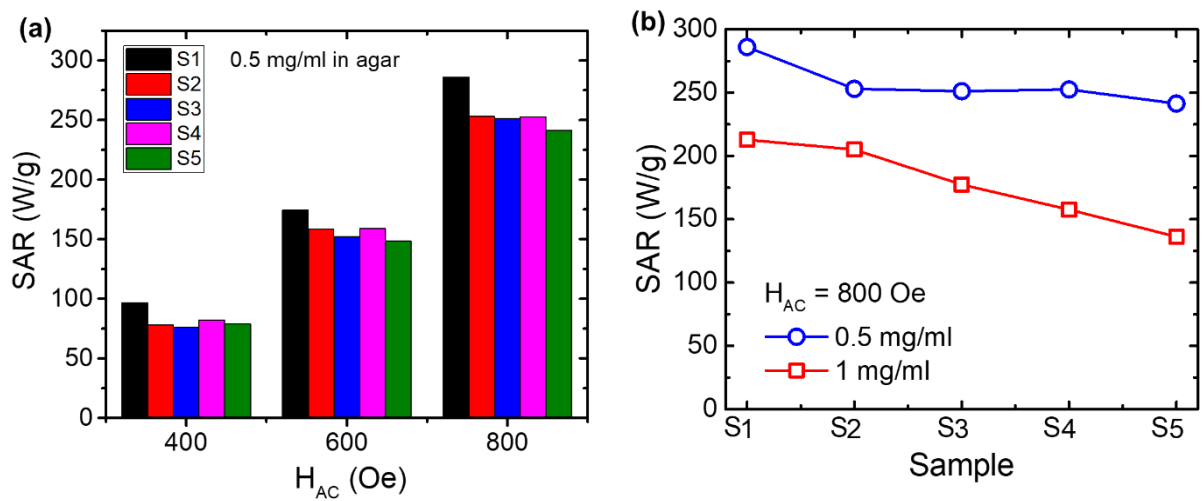

**Figure 5:** (a) SAR values of all SUPA samples in 2% agar at 0.5 mg/mL concentration in the presence of 400, 600, and 800 Oe AC fields, and (b) SAR values of all the samples in 2% agar at concentrations of 0.5 and 1 mg/mL in an 800 Oe AC field.

Figure 5a shows the SAR values for all the samples at 0.5 mg/mL concentration in 2% agar of which the $H_{AC}$ of 800 Oe showed the highest SAR values, with S1 at 286 W/g obtained using initial slope/time-rise protocol assuming a minimal to no heat gain/loss.[68] The SUPAs were initially evaluated with de-ionized water (refer to the SI, Fig. S6) which led to temperature drops with the $H_{AC}$. These drops could be related to the movement or rearrangement of the SUPAs upon application of the AC field in water. Due to inconsistencies with the temperature drops, the heating responses of SUPAs in water were not considered for the SAR calculations. Similar to the peculiar observation related to water, the concentration increments that usually lead to higher SAR values were not observed in SUPAs but rather a drop, which might be due to intensified interparticle interactions with the increased presence of SUPAs per unit volume, as shown in Fig. 5b. Moreover, we conducted comprehensive field-dependent and concentration-dependent measurements, which served to deepen our aforementioned observations (refer to the SI, Fig. S7).

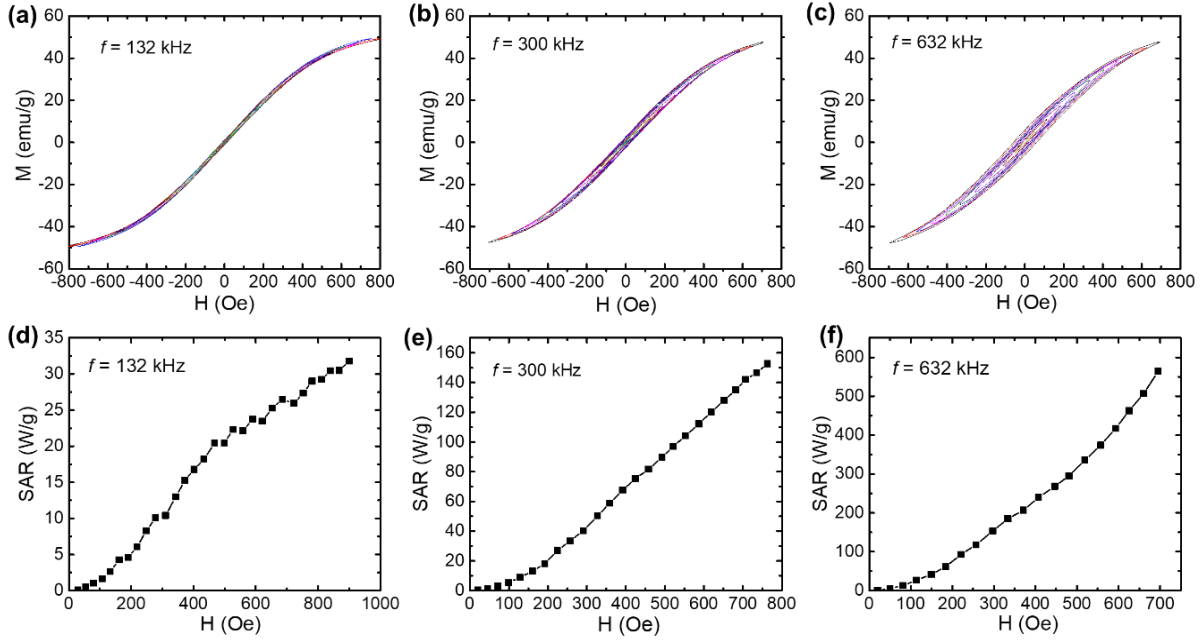

**Figure 6:** (a-c) Dynamic hysteresis loops measured at various frequencies and (d-f) the corresponding SAR vs. field curves for sample S5 at a concentration of 1.4 mg/mL.

To validate the SAR values mentioned above, we employed an additional independent method. This approach enabled us to directly measure the AC magnetic hysteresis loops and calculate the SAR values based on the area ($A$) enclosed by these loops, according to the following equation:[69]

$$SAR\left(\frac{W}{g}\right) = \frac{f}{c} \cdot A = \frac{f}{c} \cdot \oint \mu_0 M_t dH_t \qquad (5)$$

with $M_t$ being the instantaneous magnetization at time $t$, $H_t$ the sinusoidal magnetic field of frequency $f$ at time $t$, and $c$ the magnetite weight concentration in the dispersing medium. The integration was conducted over a period of the oscillating magnetic field, $T = 2\pi/f$. AC magnetic hysteresis measurements were performed at various frequencies ($f$ = 132, 300, and 638 kHz) on sample S5 possessing the largest average particle size of ~370 nm. The results obtained are displayed in Fig. 6. As illustrated, narrow minor loops were consistently obtained at low AC fields across all applied frequencies (Fig. 6a-c). However, with increasing AC field strength, both the area of the AC loops (Fig. 6a-c) and the corresponding SAR values (Fig. 6d-f) gradually increased. Notably, at 300 kHz and 800 Oe, a SAR value of approximately 150 W/g was achieved, aligning closely with the value previously estimated from calorimetric measurements (Fig. 5b).[70] It is also noteworthy from Fig. 6f that SAR reached ~600 W/g as the frequency rose to 632 kHz, indicating the potential of SUPAs for magnetic hyperthermia therapy.

Finally, we compared the SAR values for our SUPAs with those obtained from other nanostructures (spheres, cubes, rods), highlighting their superior SPM characteristics and lower concentration requirements (refer to the SI, Table S1). This comparison underscores the absence of SPM behavior in typical nanostructures as particle size increases. Notably, conventionally synthesized iron oxide nanoparticles with average sizes of 40 nm or larger exhibit ferrimagnetic (FiM) behavior, contrasting with our SUPAs, which maintain SPM characteristics despite sizes up to 400 nm. It is also noteworthy that only half the concentration of SUPAs (0.5 mg/mL vs. 1 mg/mL) is required for magnetic hyperthermia treatment compared to other candidates (Table S1). The combination of SPM properties, large and adjustable SARs, and reduced concentration requirements positions these iron oxide superparticles as compelling contenders for a wide range of biomedical applications.

## Conclusions

Polycrystalline iron oxide superparticles of particle sizes up to 400 nm, featuring crystallite sizes of 10-15 nm, exhibit superparamagnetic features at room temperature. We have demonstrated that if crystallite sizes are alike and below the SPM size threshold, SUPAs display SPM behavior at room temperature and enter a FM or FiM (blocked) state at a comparable temperature. Across all samples, a low concentration of 0.5 mg/mL consistently yielded a SAR exceeding 250 W/g at an 800 Oe AC field. The exceptional stability of superparamagnetic SUPAs in a 2% by-weight agar solution, at both 0.5 mg/mL and 1 mg/mL concentrations, underscores their potential for in vivo studies. However, further evaluations of the cytotoxicity and biocompatibility of these superparticles are needed, even though the composition and structures conform with internationally accredited safety agencies such as the FDA. The promising performance of these superparamagnetic superparticles warrants further exploration to deepen our understanding of the underlying physics behind their mechanisms.

**Experimental Methods**

**Materials.** The chemicals used for synthesis of polycrystalline superparamagnetic nanoparticles were iron(III) chloride hexahydrate (97%, Alfa Aesar), ethylene glycol (99%, Sigma Aldrich), diethylene glycol (99%, ACROS Organics), sodium acetate anhydrous (99%, ACROS Organics), sodium acetate trihydrate (99%, Fisher), sodium acrylate (97%, Sigma Aldrich), and polyethylene glycol (400) (Olin Mathieson Chemical Corporation). A 65-mL volume pressure vessel was purchased from Chem Glass. In addition, common solvents such as deionized water with a resistance of 18 MΩ-cm (Academic Milli-Q Water System, Millipore Corporation), ethanol (200 proof, Decon Labs), and acetone (99%, Oakwood) were used.

**Synthesis of iron oxide superparticles.** The polycrystalline superparamagnetic nanoparticles were synthesized by solvothermal methods in a binary solvent system containing ethylene glycol and diethylene glycol, with modifications in the technical set-up and the uses of chemical additives.[12,71–73] In this method, the use of anhydrous sodium acetate (2.17 or 3.6 g) in the synthesis was found to be effective in maintaining the small crystallite size (the size of the primary crystal) at the range from 10 to 15 nm while allowing for the fabrication of large particles with diameters larger than 160 nm.[74] Initially, a pressure vessel was charged with 1.35 g of FeCl$_3$·6H$_2$O and 20 mL of the binary solvent mixture (see below). After the complete dissolution of the iron chloride, an additional 20 mL of the binary solvent was added to anhydrous sodium acetate. The binary solvent system used for the synthesis consisted of 15/25 mL or 20/20 mL of ethylene glycol/diethylene glycol. The mixture was stirred for 30 minutes to ensure complete dissolution, followed by the addition of 1.2 mL of PEG (400) surfactant. The Teflon cap was securely fastened, and the pressure vessel was heated to 188 °C for refluxing over 5 hours. A safety shield was used to cover the synthesis setup. After synthesis, the product was cooled to room temperature and washed with ethanol combined with magnetic separation for at least 3 cycles. It is important to note that technical parameters such as stirring speed or slower heating rate can be used to alter the size of nanoparticles. Particularly, a faster stirring speed or slower ramping rate can increase the average size of nanoparticles. By adjusting the technical parameters, varying the solvent composition, and utilizing different amounts of sodium acetate additives, the size of polycrystalline iron oxide nanoparticles can be tuned from 160 to 400 nm.[74]

**Characterizations.** Scanning electron microscope (SEM) FEI Dual Beam 235 Focused Ion Beam at operating voltage 15 kV was used for imaging the nanoparticles. Samples were dissolved in ethanol and drop-cast onto a clean silicon wafer. Transmission electron microscopy (TEM) JEOL JEM-2010 was used with an accelerating voltage of 200 kV. Samples were prepared on TEM grids 300-mesh holey carbon-coated copper grids (TED Pella). X-ray diffraction data (Smart Lab, Rigaku) were collected using Cu Kα irradiation operated at 40 kV and 44 mA with a 0.01° step size was used for all samples. The crystallite size was calculated by the Scherrer formula from the diffracted peak of the (311) plane at 2θ of 35.5°. X-ray photoelectron spectroscopy (XPS) analysis was conducted on a dried sample drop-cast onto a silicon wafer using a PHI 5700 X-ray photoelectron spectrometer with monochromatic Al Kα X-rays. Calibration was performed using the C 1s peak at 284.8 eV.

The magnetic measurements were performed using a Physical Property Measurement System (PPMS) by Quantum Design, Inc., utilizing the vibrating sample magnetometer option between 10-350 K at a maximum applied magnetic field of 2 T. Calorimetric magnetic hyperthermia experiments were carried out using a 4.2 kW Ambrell Easyheat Li3542 equipment with varying AC magnetic fields (0-800 Oe) at a constant 310 kHz frequency starting at 20 $^0$C for 900 seconds with 0.5 mg/mL and 1 mg/mL nanoparticles in a 2% by weight agar solution prepared with deionized water. AC magnetometry measurements were carried out using a homemade setup to record the AC hysteresis loops.[75] The AC magnetic field amplitude was tuned between 0 and 80 mT, and 3 different frequencies were employed, 149 kHz, 300, and 638 kHz. Samples were again prepared in a 2% agar solution.

**Acknowledgments**

Work at USF was supported by the US Department of Energy, Office of Basic Energy Sciences, Division of Materials Science and Engineering under Grant No. DE-FG02-07ER46438. J.A. acknowledges financial support from Spanish MCIN/AEI under Project PID2020-115704RB-C3., T.R. Lee thanks financial support for the research at UH from the Air Force Office of Scientific Research (AFOSR FA9550-23-1-0581; 23RT0567) and the Robert A. Welch Foundation (Grant Nos. E-1320 and V-E-0001. H.S. acknowledges support from US National Science Foundation through Grant No. NSF DMR-2327667 and also support from Alexander von Humboldt foundation for a Humboldt Research Award. M.H. Phan acknowledges the Programa de Profesores Visitantes 2023-2024 offered by the Vice-Rector of Research and Knowledge Transfer at the University of Cantabria, Spain for supporting his visiting professorship.

Table of Contents

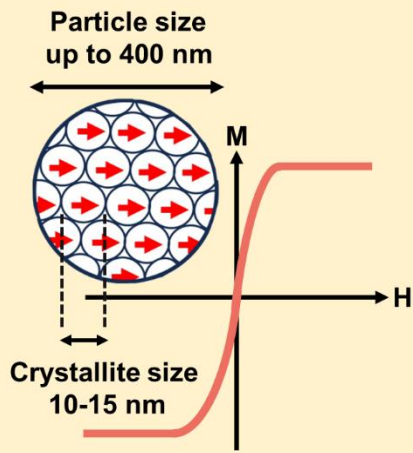
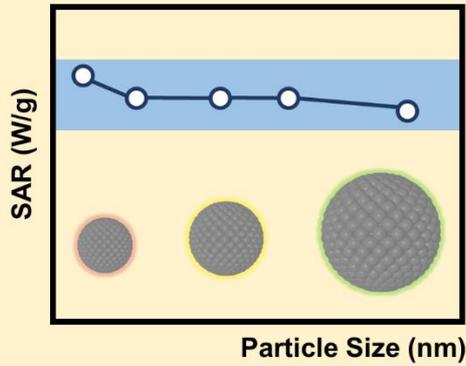
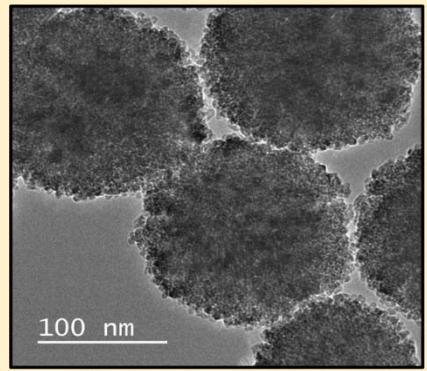

# Supporting Information

**Superparamagnetic Superparticles for Magnetic Hyperthermia Therapy: Overcoming the Particle Size Limit**

Supun B. Attanayake[1,#], Minh Dang Nguyen[2,#], Amit Chanda[1], Javier Alonso[3], Iñaki Orue[4], T. Randall Lee[2,*], Hariharan Srikanth[1], and Manh-Huong Phan[1,*]

[1]Department of Physics, University of South Florida, Tampa, FL 33620, USA

[2]Department of Chemistry and the Texas Center for Superconductivity, University of Houston, 4800 Calhoun Road, Houston, TX 77204-5003, USA

[3]Department of CITIMAC, Universidad de Cantabria, Santander 39005, Spain

[4]SGIker Medidas Magnéticas, Universidad del País Vasco, Leioa 48940, Spain

*Corresponding authors: phanm@usf.edu (M.H.P.); trlee@uh.edu (T.R.L.)

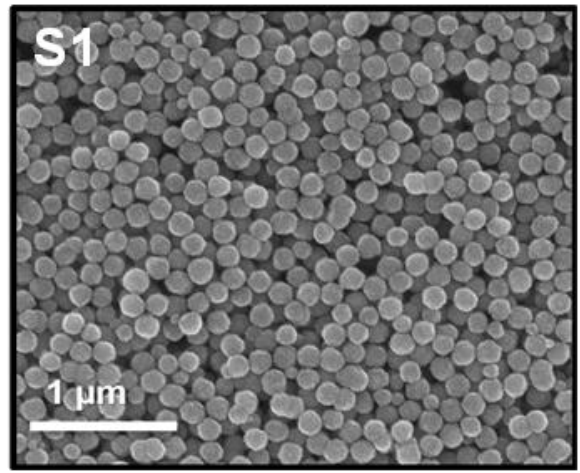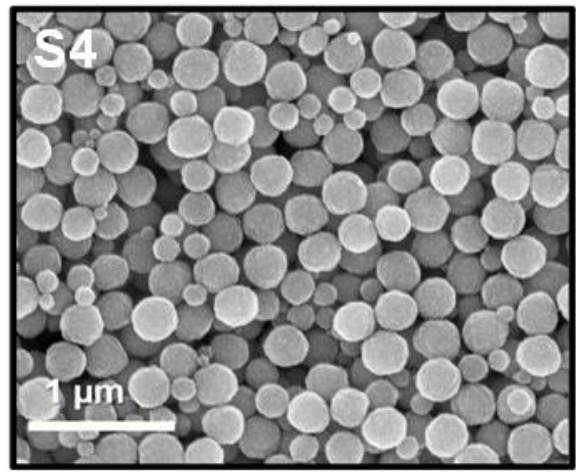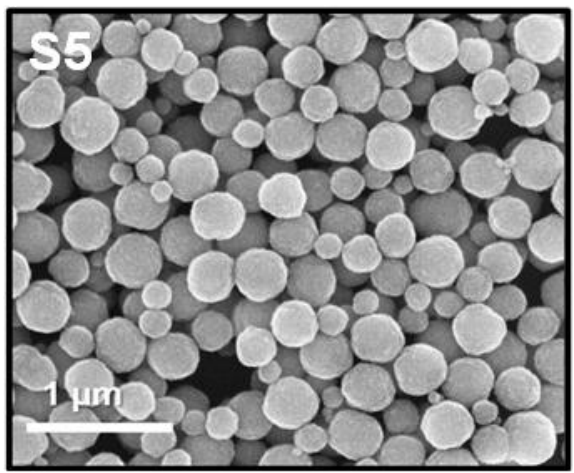

**Figure S1.** SEM images of SUPA samples S1, S4, and S5.

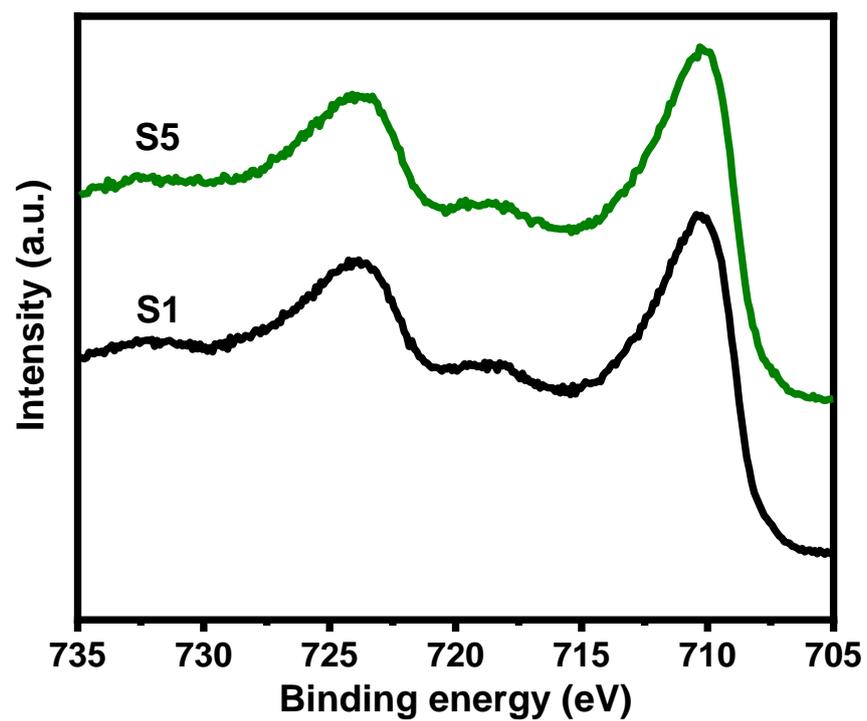

**Figure S2.** XPS spectra of SUPA samples S1 and S5. A detailed analysis of these spectra confirms the presence of $Fe_3O_4$ as a major phase and $\gamma\text{-}Fe_2O_3$ as a minor phase.

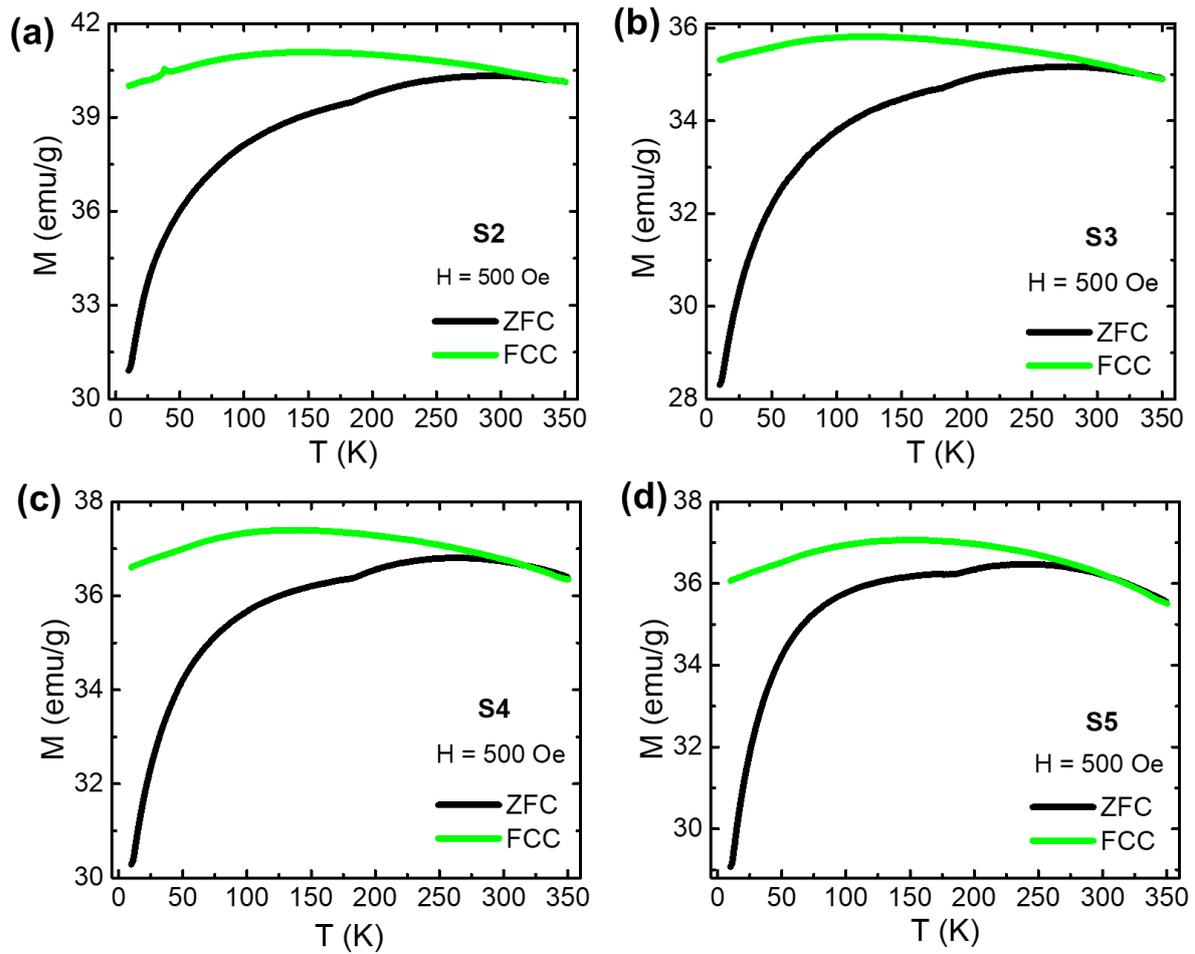

**Figure S3.** The temperature-dependent magnetization *M*(*T*) curves under ZFC and FC protocols for samples S2, S3, S4, and S5. The observation of consistent magnetic behavior suggests that the magnetic properties of the superparticles are predominantly determined by the size of the nanocrystals composing SUPAs.

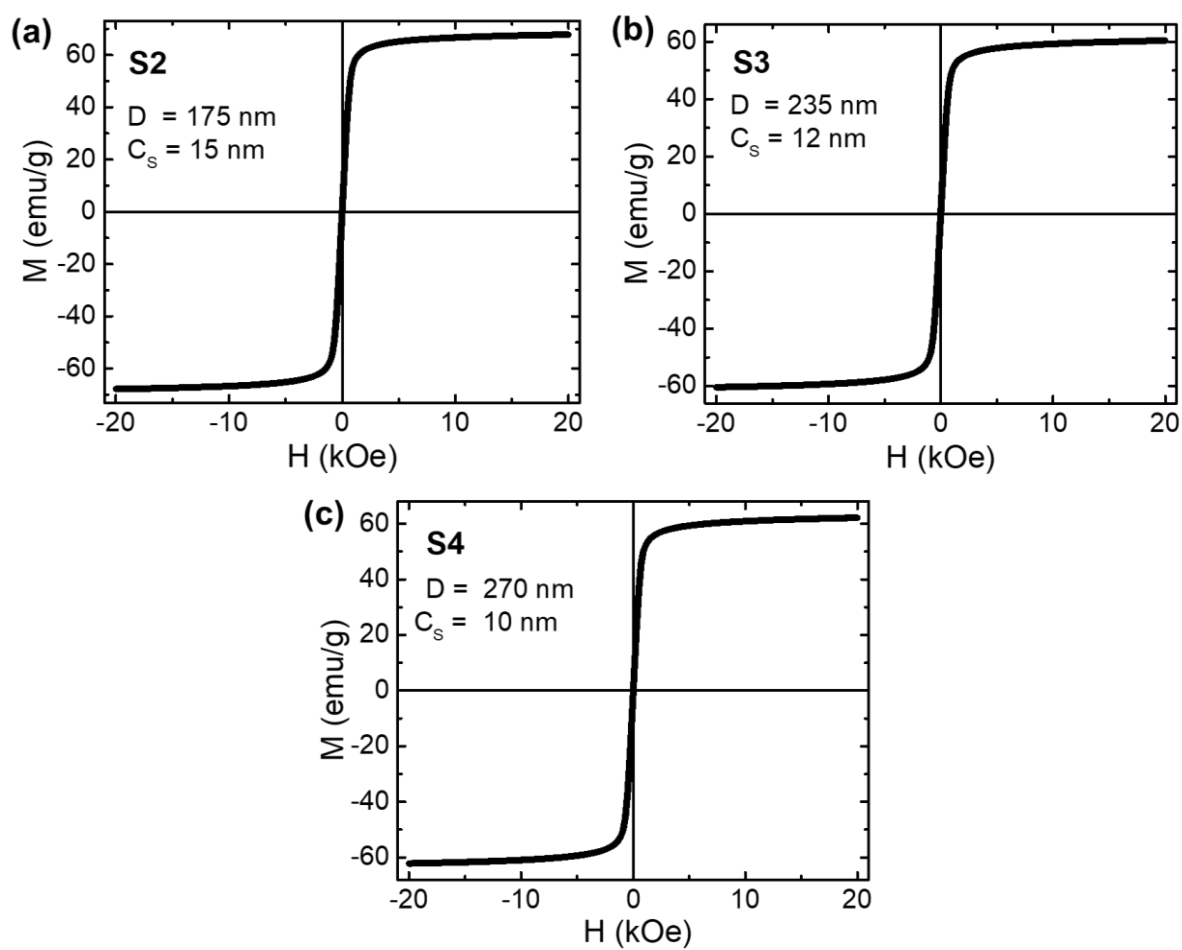

**Figure S4.** The magnetic field-dependent magnetization *M*(*H*) curves, measured at 300 K, for samples S2, S3, and S4. The absence of magnetic hysteresis in these curves confirms the superparamagnetic characteristics of the samples.

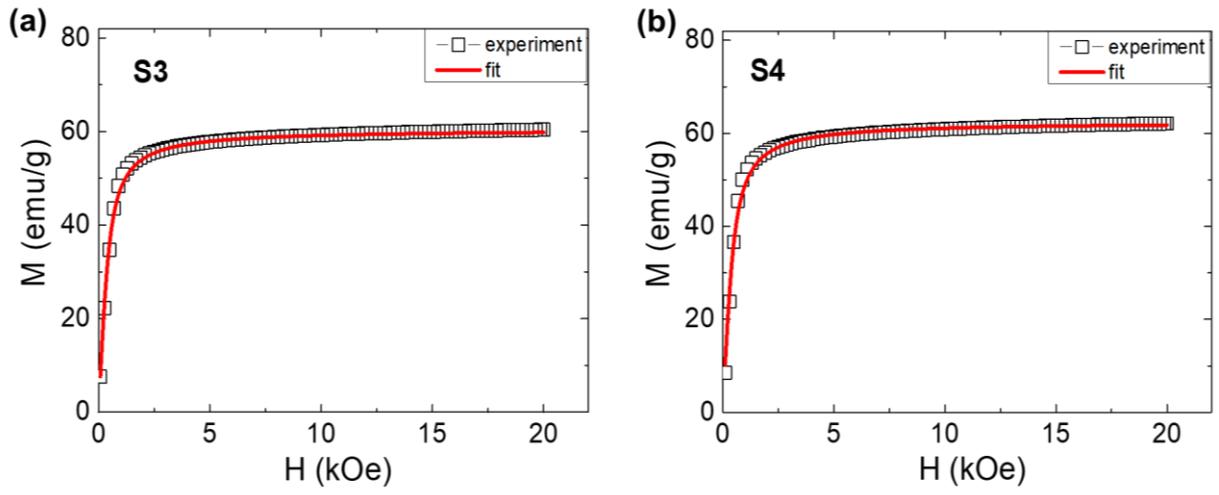

**Figure S5.** The *M*(*H*) results obtained from the experiment and fitting of the Langevin Function (Eq. (1)) for samples (a) S3 and (b) S4.

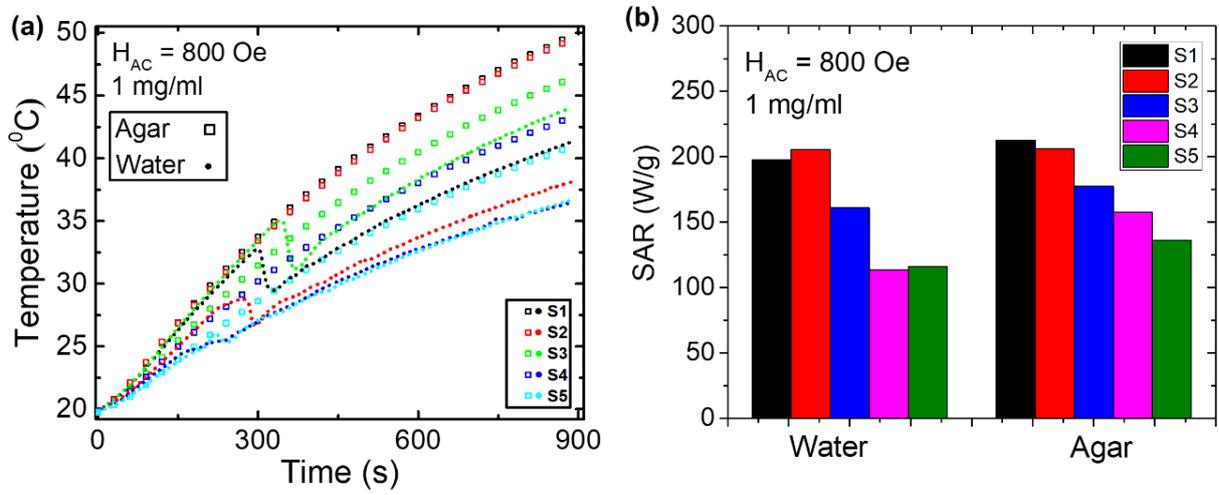

**Figure S6.** (a) Heating curves for the samples at a concentration of 1 mg/mL and (b) their corresponding SAR values measured at an 800 Oe AC field and a frequency of 310 kHz in both water and agar.

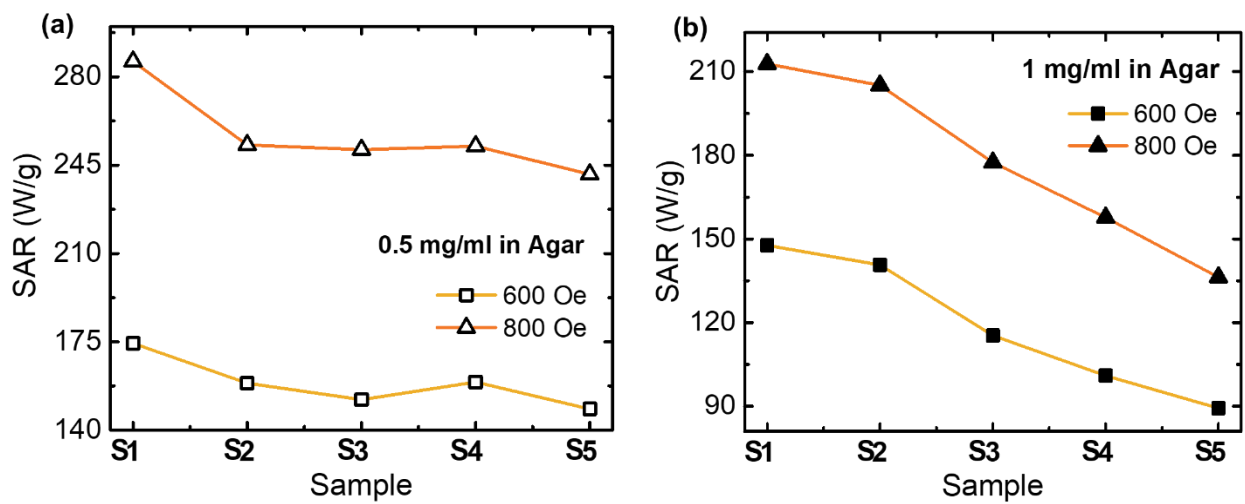

**Figure S7.** SAR values of all samples measured at 600 and 800 Oe AC fields for concentrations of (a) 0.5 and (b) 1.0 mg/mL in agar.

**Table S1:** Particle size ($D$), concentration, saturation magnetization ($M_s$), coercivity ($H_c$), and SAR values of various SPM nanosystems.

| Sample | $D$ (nm) | Concentration (mg/mL) | SAR in Agar at 800 Oe (W/g) | $M_s$ at 300 K (emu/g) | $H_c$ (Oe) | Ref. |
|---|---|---|---|---|---|---|
| Nanospheres | 25 | 1 | 160 | 68 | ~0 | [1] |
| Nanocubes | 19 | 1 | 280 | 75 | ~0 | |
| Nanorods | 41 | 1 | 350 | 86 | ~0 | [2] |
| SUPA (S1) | 160 | 0.5 | 286 | 66 | ~0 | This work |
| SUPA (S5) | 375 | 0.5 | 241 | 61 | ~0 | |